\documentstyle[12pt,subeqn]{article}
\begin{document}

\begin{titlepage}
\begin{center}
\huge{The Dirac Equation Is Separable On The Dyon Black Hole Metric} \\
\vspace{1 cm}
\large{\.{I}brahim Semiz} \footnote{e-mail(bitnet): semiz@yalehep} \\
\vspace {1 cm}
\normalsize{Department of Physics, Yale University\\
217 Prospect St. \\ New Haven, CT 06511\\}
\vspace { 1 cm }
\begin{abstract}
Using the tetrad formalism, we carry out the separation
of variables for the massive complex
Dirac equation in the gravitational and electromagnetic
field of a \mbox{four-parameter} \mbox{(mass,} angular momentum, electric and
magnetic \mbox{charges)} black hole.
\end{abstract}
\vspace {2 cm}
PACS Numbers: 04.40.+c, 97.60.Lf
\end{center}
\end{titlepage}

\section{Introduction}

\hspace{5 mm}
In studies of the behavior of a matter field on a black hole background,
e.g. the possible evolution of the black hole by exchanging energy,
charge and angular momentum with the field; one would like to know if
the variables can be separated in the relevant field equation.
Obviously, the study becomes much easier, if the system of partial
differential equations can be reduced to a system of ordinary differential
equations. This paper concentrates on \mbox{spin-$\frac{1}{2}$} (Dirac) field.

The \mbox{``no-hair''} theorem in General Relativity asserts that the
metrics of stationary black holes can be described uniquely by three
parameters : \mbox{Mass $M$,} \mbox{charge $Q_{e}$} \mbox{(assuming} the
absence of magnetic \mbox{charges),} and angular momentum per unit mass,
$a$. Therefore, when one writes a matter field equation on a black hole
background, these parameters become parameters of the field equation.

The Dirac equation and its separability properties on black hole metrics
have been investigated in increasingly complicated contexts. The early
pioneering work was done by Brill and Wheeler \cite{brill-whe}, who
separated the neutrino field equation on the Schwarzschild metric.
Teukolsky \cite{teukolsky} separated a master equation for massles
spin 0, 1 and 2 and noted that the separated equations for the massless
\mbox{spin-$\frac{1}{2}$} have the same form, and therefore could be
incorporated into the master equation. Unruh \cite{unruh} showed the
\mbox{spin-$\frac{1}{2}$} result independently. The unexpected result
that the massive Dirac Equation was also separable came from
Chandrasekhar \cite{chandra.paper}. The Chandrasekhar result was
generalized to the \mbox{Kerr-Newman} \mbox{($Q,a,M$)} case by Page \cite{page}
and Toop \cite{toop}.

However, black holes could also have magnetic charge, if such existed.
Such a black hole would acquire an additional label $Q_{m}$ for the
magnetic charge. The interest in this possibility has grown since
magnetic monopoles have been found to be required in various extensions
of the standard model of particle physics. Dudley and Finley
\cite{dudley-finley}, carried out the separation of variables for all
{\em real, massless, single spin} field equations with $s=0$,
$\frac{1}{2}$, $1$, $2$ on the \mbox{seven-parameter} class of
\mbox{Petrov type-D} solutions of \mbox{Einstein-Maxwell} equations found by
Plebanski and Demianski \cite{plebanski-dem}. The seven parameters
include $M$, $a$, $Q_{e}$ and $Q_{m}$ (The coordinates are not of
\mbox{Boyer-Lindquist} type). Since the fields are real, these results
apply only to neutral particles.
In this paper, we use the tetrad formalism to show that the Dirac
equation for massive, charged fermions remains separable when magnetic
charge is added to the black hole, i.e. in the dyonic \mbox{Kerr-Newman} case;
and we present the separated radial and angular equations.
We anticipate using this separability in a thought experiment to test
the cosmic censorship conjecture by considering a massive, charged Dirac
field on a dyonic black hole, i.e. the \mbox{Dirac-field} analogue of work done
in \cite{semiz1}.

\section{The Tetrad Formalism}

\hspace{5 mm}
The equivalence principle tells us that the laws of Physics, as written
in Minkowski space, are also valid in a freely falling frame on a curved space,
i.e. a local Lorentz coordinate system (LLCS). This principle can be utilized
to construct the {\em tetrad formalism}\footnote{See, e.g. Weinberg \cite{wbg},
Section 12.5, Birrell \& Davies \cite{bir-dav}, Section 3.8, Chandrasekhar
\cite{chandra.book}, \S 7.}, a prescription for writing the laws of
Physics on curved spaces. We erect at each point in spacetime, four vectors
$V_{a}^{\;\;\mu}$, the ``basis vectors of a LLCS". Here, the latin
indices from the early part of the alphabet label the coordinates in the LLCS,
and the greek indices, the spacetime
coordinates. The set of four $V_{a}$'s, the {\em tetrad} or
{\em vierbein}, is usually chosen to be orthonormal:
\begin{equation}
g_{\mu \nu} V_{a}^{\;\mu} V_{b}^{\;\nu}
= \eta_{a b} \; .     \label{eq:normalisation}
\end{equation}
The LLCS indices are raised and lowered with the Minkowski tensor
$\eta_{a b}$ (we use \mbox{$\eta_{a b} =$ {\sl diag}(-1,1,1,1))}
and spacetime indices with the metric $g_{\mu \nu}$;
the tetrad satisfies
\begin{equation}
\eta_{{a} {b}} V^{{a}}_{\;\;\mu} V^{{b}}_{\;\;\nu}
= g_{\mu \nu}          \label{eq:decomposition}
\end{equation}
We can write down components of physical fields in the LLCS:
\begin{equation}
A^{{a} {b} \cdots}_{{c} {d} \cdots}
= V^{{a}}_{\;\;\mu} V^{{b}}_{\;\;\nu} V_{{c}}^{\;\sigma}
V_{{d}}^{\;\eta} \cdots
A^{{\mu} {\nu} \cdots}_{{\sigma} {\eta} \cdots}    \label{eq:LLCSfields}
\end{equation}
These objects $A^{{a} {b} \cdots}_{{c} {d} \cdots}$ are scalars
with respect to spacetime operations, and tensors in
the LLCS, the number of the covariant and contravariant indices being the same
as for the original object
$A^{{\mu} {\nu} \cdots}_{{\sigma} {\eta} \cdots}$

In terms of these components, we want to write down laws of Physics in the
LLCS,
in their Minkowskian form. We also want the laws to be invariant under local
Lorentz transformations, i.e. different tetrad choices. This leads us to define
the \mbox{coordinate-scalar}, \mbox{Lorentz-vector} `derivative'
\begin{equation}
{\cal D}_{{a}} = V_{{a}}^{\;\mu} \partial_{\mu}
+ \frac{1}{2} V_{{a}}^{\;\mu} V_{{b}}^{\;\nu}
  V_{{c} \nu ; \mu} \Sigma^{{b} {c}}          \label{eq:LLCSderv}
\end{equation}
where $\Sigma^{\alpha \beta}$ are generators of the Lorentz group in the
representation associated with the field ${\cal D}_{{a}}$ is acting on;
and the semicolon denotes the \mbox{(metric-)covariant}
derivative\footnote{Weinberg \cite{wbg} and Birell \& Davies \cite{bir-dav}
give --incorrectly-- an ordinary derivative here.}.

Now the prescription is: We take the Minkowskian form of the relevant equation,
replace each term by its LLCS equivalent via eqn.~(\ref{eq:LLCSfields}) and
each derivative by its LLCS equivalent via eqn.~(\ref{eq:LLCSderv}).

\section{The Dirac Equation on the dyonic black hole metric}

\hspace{5 mm}
The Dirac Equation in flat space with a background electromagnetic field is
\begin{equation}
i \gamma^{a} (\partial_{a} + i e A_{a}) \Psi = \mu \; \Psi
                                                     \label{eq:mkdirac}
\end{equation}
where $\Psi$ is a \mbox{four-component} spinor, $\mu$ the mass  and $e$ the
charge of the field quantum, $A$ the \mbox{vector-potential} of the
electromagnetic field and the constant \mbox{$\gamma$-matrices} satisfy
\begin{equation}
\{\gamma^{a}, \gamma^{b} \} = -2 \; \eta^{ab}.         \label{eq:acomm}
\end{equation}
We will be using \mbox{$\gamma$-matrices} of the form
\begin{equation}
\gamma^{{0}} = \left( \begin{array}{cc}
                                       0 &  \sigma^{{0}} \\
                                       \sigma^{{0}} & 0 \end{array}
                   \right), \hspace {0.5 cm}
\gamma^{{i}} = \left( \begin{array}{cc}
                                       0 & -\sigma^{{i}} \\
                                       \sigma^{{i}} & 0 \end{array}
                   \right)
\end{equation}
where the \mbox{$\sigma$-matrices} are the identity matrix and the Pauli
matrice
   s:
\begin{equation}
\sigma^{{0}} = \left( \begin{array}{ll}1 & 0 \\
                                 0 & 1 \end{array} \right), \hspace {0.5 cm}
\sigma^{{1}} = \left( \begin{array}{ll}0 & 1 \\
                                 1 & 0 \end{array} \right), \hspace {0.5 cm}
\sigma^{{2}} = \left( \begin{array}{lr}0 & -i \\
                                 i & 0 \end{array} \right), \hspace {0.5 cm}
\sigma^{{3}} = \left( \begin{array}{lr}1 & 0 \\
                                 0 & -1 \end{array} \right)
\end{equation}
For a \mbox{spin-$\frac{1}{2}$} field, the \mbox{$\Sigma$-matrices} are given
by
\begin{equation}
\Sigma^{{b} {c}} = - \frac{1}{4} [\gamma^{{b}}, \gamma^{{c}}] .
\end{equation}

To get the Dirac Equation in curved space, we follow the prescription and make
the replacements (\ref{eq:LLCSfields},\ref{eq:LLCSderv}):
\begin{equation}
i [\gamma^{a}]_{k}^{\;\;j} V_{{a}}^{\;\mu} \partial_{\mu} \Psi_{j}
+ \frac{1}{2} i [\gamma^{a}]_{k}^{\;\;l} V_{{a}}^{\;\mu}
         V_{{b}}^{\;\nu}   V_{{c}\nu ; \mu} \;
         [\Sigma^{{b} {c}}]_{l}^{\;\;j} \Psi_{j}
- e [\gamma^{a}]_{k}^{\;\;j} V_{{a}}^{\;\mu} A_{\mu}) \Psi_{j}
= \mu \; \Psi_{k}                                        \label{eq:curvdirac}
\end{equation}
where we have written out the spinor indices $j,k,l$ explicitly.

The metric of the dyonic black hole is\footnote{For derivations in
\mbox{Boyer-Lindquist} coordinates, see \cite{semiz1} and \cite{kasuya}.}
\begin{subequations}
\begin{eqnarray}
ds^{2} = g_{\mu\nu}dx^{\mu}dx^{\nu}
       & = & \frac{a^{2}\sin^{2}\!\theta-\Delta}{\rho^{2}}dt^{2}
              +\frac{(r^{2}+a^{2})^{2}-\Delta a^{2}\sin^{2}\!\theta}
                    {\rho^{2}}    \sin^{2}\!\theta  \; d\phi^{2}
                                                        \nonumber \\
       & + & 2\frac{\Delta-(r^{2}+a^{2})}{\rho^{2}}a\sin^{2}\!\theta \;
                                                             dt d\phi
                    +\frac{\rho^{2}}{\Delta}dr^{2}+\rho^{2}d\theta^{2}
\end{eqnarray}
where
\begin{eqnarray}
\rho^{2} & = & r^{2}+a^{2}\cos^{2}\!\theta\\
\Delta   & = & r^{2}-2Mr+a^{2}+Q_{e}^{\:\:2}+Q_{m}^{\:\:2} \label{eq:Delta}
\end{eqnarray}
\end{subequations}
with the vector potential
\begin{subequations}
\begin{eqnarray}
A_{t} & = & -Q_{e}\frac{r}{\rho^{2}}+Q_{m}\frac{a\cos\! \theta}{\rho^{2}}\\
A_{r} & = & A_{\theta}=0\\
A_{\phi} & = & Q_{e}\frac{ar\sin^{2}\!\theta}{\rho^{2}}
+Q_{m} [\pm1-\cos\! \theta\frac{r^{2}+a^{2}}{\rho^{2}}] .
\end{eqnarray}
\end{subequations}
and we are using \mbox{Boyer-Lindquist} coordinates.

 Of course, the vector potential is unique only up to a gauge
transformation,
and the magnetic part of $A_{\phi}$ contains a string singularity.
The two signs in that term correspond to the two gauges we will be
using. The
\mbox{upper-sign} term puts the string along the negative
\mbox{$z$-axis}
\mbox{$(\theta=\pi)$} and will be used when
\mbox{$0\leq \theta \leq \pi/2$,}
the \mbox{lower-sign} term puts it along the positive \mbox{$z$-axis}
\mbox{$(\theta=0)$}
and will be used when \mbox{$\pi/2 < \theta \leq \pi$.} Therefore, the
wavefunction is also gauge transformed across the equator, and picks up
a
factor of $e^{2ieQ_{m}\phi}$ passing from north to south. This matching
of
boundary conditions ensures that the problem can be expressed
meaningfully
without strings of diverging vector potential. Such a wavefunction is
called
a section \cite{wu-yang}.

We will derive separated equations first using a simple tetrad that we
call `canonical'. For purposes of comparison at the appropriate
limit with Chandrasekhar \cite{chandra.paper}, we will repeat the calculation
using a more complicated tetrad, calling it \mbox{`Kinnersley-type'}. The
labels $(C)$ and $(K)$ will be used
for variables whenever it is not obvious which case they belong to or for
emphasis.
\subsection{The Canonical Tetrad}
\hspace{5 mm}
We use the generalization of the simple `canonical' tetrad emphasized by
Carter, relabeling the vectors to be able
to compare with Carter \& McLenaghan \cite{cart-mcl.MG2} in the
limit $Q_{m} = 0$:
\begin{subequations}
\begin{eqnarray}
V^{(C)0} & = & \frac{\delta}{\rho} (dt - a \sin^{2}\! \theta \; d\phi) \\
V^{(C)1}
    & = & \frac{\sin\! \theta}{\rho} [-a \; dt + (r^{2}+a^{2}) \; d\phi)] \\
V^{(C)2} & = & \rho \; d\theta \\
V^{(C)3} & = & \frac{\rho}{\delta} \; dr
\end{eqnarray}
\end{subequations}
where $\delta = \sqrt{\Delta}$ and $\rho$ was defined above via its square.

Making all the necessary substitutions into eq.~(\ref{eq:curvdirac}), we
get the four coupled components of the Dirac equation\footnote{The calculations
for going from eqn.~(\ref{eq:curvdirac}) to
eqns.~(\ref{eq:expdirac1.c}-\ref{eq:expdirac4.c}) or
eqns.~(\ref{eq:expdirac1.k}-\ref{eq:expdirac4.k}) have been performed using
the {\em Mathematica} symbolic mathematics software, and the `tetrad' package
\cite{tetrad.m}.} on the dyon black hole metric:
\begin{subequations}
\begin{eqnarray}
\lefteqn{ \left[ \frac{(r^{2}+a^{2})}{\rho\delta} \partial_{t}
      -\frac{\delta}{\rho} \partial_{r}
      +\frac{a}{\rho\delta} \partial_{\phi}
      - \frac{\delta}{2\rho\bar{\rho}^{*}} - \frac{(r-M)}{2\rho\delta}
      + ie \frac{-Q_{e}r \pm a Q_{m}}{\rho\delta} \right]
        \Psi_{2}^{(C)} }
      \hspace{1 cm}     \nonumber \\
& & + \left[ -\frac{a \sin\! \theta}{\rho} \partial_{t}
         + \frac{i}{\rho} \partial_{\theta}
         - \frac{1}{\rho \sin\! \theta} \partial_{\phi}
         - \frac{a \sin\! \theta}{2\rho\bar{\rho}^{*}}
         + \frac{i}{2\rho} \cot\! \theta
   - i e Q_{m} \frac{\pm 1 - \cos\! \theta}{\rho \sin\! \theta} \right]
         \Psi_{3}^{(C)}
= - i \mu \Psi_{0}^{(C)}  \hspace{1 cm} \label{eq:expdirac1.c}  \\
\lefteqn{ \left[  - \frac{a \sin\! \theta}{\rho} \partial_{t}
         - \frac{i}{\rho} \partial_{\theta}
         - \frac{1}{\rho \sin\! \theta} \partial_{\phi}
         + \frac{a \sin\! \theta}{2\rho\bar{\rho}^{*}}
         - \frac{i}{2\rho} \cot\! \theta
   - i e Q_{m} \frac{\pm 1 - \cos\! \theta}{\rho \sin\! \theta} \right]
          \Psi_{2}^{(C)}
               } \hspace{1 cm} \nonumber \\
& & + \left[ \frac{(r^{2}+a^{2})}{\rho\delta} \partial_{t}
      +\frac{\delta}{\rho} \partial{r}
      +\frac{a}{\rho\delta} \partial{\phi}
      + \frac{\delta}{2\rho\bar{\rho}^{*}} + \frac{(r-M)}{2\rho\delta}
      + ie \frac{-Q_{e}r \pm a Q_{m}}{\rho\delta} \right]
                                  \Psi_{3}^{(C)}
= - i \mu \Psi_{1}^{(C)}   \hspace{1 cm}  \\
\lefteqn{ \left[ \frac{(r^{2}+a^{2})}{\rho\delta} \partial_{t}
      +\frac{\delta}{\rho} \partial{r}
      +\frac{a}{\rho\delta} \partial{\phi}
      + \frac{\delta}{2\rho\bar{\rho}} + \frac{(r-M)}{2\rho\delta}
      + ie \frac{-Q_{e}r \pm a Q_{m}}{\rho\delta} \right]
                                    \Psi_{0}^{(C)}
      } \hspace{1 cm} \nonumber \\
& & + \left[ \frac{a \sin\! \theta}{\rho} \partial_{t}
         - \frac{i}{\rho} \partial_{\theta}
         + \frac{1}{\rho \sin\! \theta} \partial_{\phi}
         - \frac{a \sin\! \theta}{2\rho\bar{\rho}}
         - \frac{i}{2\rho} \cot\! \theta
   + i e Q_{m} \frac{\pm 1 - \cos\! \theta}{\rho \sin\! \theta} \right]
                                                \Psi_{1}^{(C)}
= - i \mu \Psi_{2}^{(C)}     \hspace{1 cm} \\
\lefteqn{\left[ \frac{a \sin\! \theta}{\rho} \partial_{t}
         +\frac{i}{\rho} \partial_{\theta}
         +\frac{1}{\rho \sin\! \theta} \partial_{\phi}
         + \frac{a \sin\! \theta}{2\rho\bar{\rho}}
         + \frac{i}{2\rho} \cot\! \theta
    + ie Q_{m} \frac{\pm 1 - \cos\! \theta}{\rho \sin\! \theta} \right]
                                              \Psi_{0}^{(C)}
         } \hspace{1 cm} \nonumber \\
& & + \left[\frac{(r^{2}+a^{2})}{\rho\delta} \partial_{t}
      -\frac{\delta}{\rho} \partial{r}
      + \frac{a}{\rho\delta} \partial{\phi}
      - \frac{\delta}{2\rho\bar{\rho}} - \frac{(r-M)}{2\rho\delta}
      + ie \frac{-Q_{e}r \pm a Q_{m}}{\rho\delta} \right]
                                     \Psi_{1}^{(C)}
= - i \mu \Psi_{3}^{(C)}  \hspace{1 cm} \label{eq:expdirac4.c}
\end{eqnarray}
\end{subequations}
where
\begin{displaymath}
\bar{\rho} = r + i a \cos\! \theta,  \;\;\; \bar{\rho}^{*} = r - i a \cos\!
\theta,
\;\;\; \bar{\rho} \bar{\rho}^{*} = \rho^{2}
\end{displaymath}
\subsection{The `Kinnersley-type' Tetrad}
\hspace{5 mm}
For sake of comparison with Chandrasekhar \cite{chandra.paper}, in the limit
$Q_{e} = 0$, $Q_{m} = 0$, we write down the Dirac Equation using the
generalization of the \mbox{Kinnersley-type} tetrad corresponding to the null
tetrad used by him:
\begin{subequations}
\begin{eqnarray}
V^{(K)0} & = & \frac{1}{\sqrt{2}\,}[ (1 + \frac{\Delta}{2 \rho^{2}}) \; dt
                        + (\frac{1}{2} - \frac{\rho^{2}}{\Delta}) \; dr
          - (1 + \frac{\Delta}{2 \rho^{2}}) a  \sin^{2}\!\theta \; d\phi]
                  \\
V^{(K)1}
 & = & - \frac{a^{2} \cos\!\theta \sin\!\theta}{\rho^{2}} \; dt + r \; dr
     + \frac{a (a^{2}+r^{2}) \cos\!\theta \sin\!\theta}{\rho^{2}} \; d\phi
                                                          \\
V^{(K)2} & = & \frac{a r \sin\!\theta}{\rho^{2}}\; dt
          + a \cos\!\theta \; d\theta
             - \frac{r (a^{2}+r^{2}) \sin\!\theta}{\rho^{2}} \; d\phi \\
V^{(K)3} & = & \frac{1}{\sqrt{2}\,}[(-1 + \frac{\Delta}{2 \rho^{2}}) \; dt
                        + (\frac{1}{2} + \frac{\rho^{2}}{\Delta}) \; dr
          + (1 - \frac{\Delta}{2 \rho^{2}}) a  \sin^{2}\!\theta \; d\phi]
\end{eqnarray}
\end{subequations}

Once again, making necessary get four coupled components of the Dirac
equation on the dyon black hole metric;
for the \mbox{Kinnersley-type} tetrad:
\begin{subequations}
\begin{eqnarray}
\lefteqn{ \left[\frac{(r^{2}+a^{2})} {\sqrt{2}\, \rho^{2}}  \partial_{t}
        - \frac{\Delta} {\sqrt{2}\, \rho^{2}} \partial_{r}
        + \frac{a} {\sqrt{2}\, \rho^{2}} \partial_{\phi}
        - \frac{(r-M)} {\sqrt{2}\, \rho^{2}}
        + \frac{ie} {\sqrt{2}\,}
                           \left( \frac{\pm a Q_{m} -Q_{e} r} {\rho^{2}}
\right)
  \right] \Psi_{2}^{(K)} } \nonumber \\
& & \hspace{-.5 cm}+ \left[- \frac{i a \sin\! \theta}{\bar{\rho}}  \partial_{t}
        - \frac{1} {\bar{\rho}} \partial_{\theta}
        - \frac{i} {\bar{\rho} \sin\! \theta} \partial_{\phi}
        - \left(\frac{i a \sin\! \theta} {\rho^{2}}
                                  + \frac{1} {2 \bar{\rho}} \cot \theta \right)
       + e Q_{m} \left( \frac{\pm 1 - \cos\! \theta} {\bar{\rho} \sin\! \theta}
\right)
  \right] \Psi_{3}^{(K)}
      = - i \mu \Psi_{0}^{(K)} \hspace{1 cm}
                                   \label{eq:expdirac1.k} \\
\lefteqn{
  \left[\frac{i a \sin\! \theta}{\bar{\rho}^{*}}  \partial_{t}
        - \frac{1} {\bar{\rho}^{*}} \partial_{\theta}
        + \frac{i} {\bar{\rho}^{*} \sin\! \theta} \partial_{\phi}
        - \frac{1} {2 \bar{\rho}^{*}} \cot \theta
        - e Q_{m} \left( \frac{\pm 1 - \cos\! \theta}
                                          {\bar{\rho}^{*} \sin\! \theta}
\right)
  \right] \Psi_{2}^{(K)} } \nonumber \\
& & 
+ \left[\sqrt{2}\, \frac{(r^{2}+a^{2})} {\Delta} \partial_{t}
        + \sqrt{2}\, \partial_{r}
        + \sqrt{2}\, \frac{a} {\Delta} \partial_{\phi}
        + \sqrt{2}\, \frac{1} {\bar{\rho}^{*}}
        + ie \sqrt{2}\, \left( \frac{\pm a Q_{m} -Q_{e} r} {\Delta} \right)
  \right] \Psi_{3}^{(K)}
       = - i \mu \Psi_{1}^{(K)} \\
\lefteqn{
\left[\sqrt{2}\, \frac{(r^{2}+a^{2})} {\Delta} \partial_{t}
        + \sqrt{2}\, \partial_{r}
        + \sqrt{2}\, \frac{a} {\Delta} \partial_{\phi}
        + \sqrt{2}\, \frac{1} {\bar{\rho}}
        + ie \sqrt{2}\, \left( \frac{\pm a Q_{m} -Q_{e} r} {\Delta} \right)
  \right] \Psi_{0}^{(K)} } \nonumber \\
& & \hspace{2 cm} + \left[\frac{i a \sin\! \theta}{\bar{\rho}}  \partial_{t}
        + \frac{1} {\bar{\rho}} \partial_{\theta}
        + \frac{i} {\bar{\rho} \sin\! \theta} \partial_{\phi}
        + \frac{1} {2 \bar{\rho}} \cot \theta
        - e Q_{m} \left( \frac{\pm 1 - \cos\! \theta}
                                              {\bar{\rho} \sin\! \theta}
\right)
  \right] \Psi_{1}^{(K)}
       = - i \mu \Psi_{2}^{(K)} \\
\lefteqn{
  \left[- \frac{i a \sin\! \theta}{\bar{\rho}^{*}}  \partial_{t}
        + \frac{1} {\bar{\rho}^{*}} \partial_{\theta}
        - \frac{i} {\bar{\rho}^{*} \sin\! \theta} \partial_{\phi}
        + \left(- \frac{i a \sin\! \theta} {\rho^{2}}
                                + \frac{1}{2 \bar{\rho}^{*}} \cot \theta
\right)
        + e Q_{m} \left( \frac{\pm 1 - \cos\! \theta}
                                          {\bar{\rho}^{*} \sin\! \theta}
\right)
  \right] \Psi_{0}^{(K)} } \nonumber \\
& & \hspace{.5 cm}
+ \left[\frac{(r^{2}+a^{2})} {\sqrt{2}\, \rho^{2}} \partial_{t}
        - \frac{\Delta} {\sqrt{2}\, \rho^{2}} \partial_{r}
        + \frac{a} {\sqrt{2}\, \rho^{2}} \partial_{\phi}
        - \frac{r-M} {\sqrt{2}\, \rho^{2}}
        + \frac{ie} {\sqrt{2}}\, \left( \frac{\pm a Q_{m} -Q_{e} r} {\rho^{2}}
                                                                        \right)
  \right] \Psi_{1}^{(K)}
       = - i \mu \Psi_{3}^{(K)}            \label{eq:expdirac4.k}
\end{eqnarray}
\end{subequations}

\section{The Separation}
\hspace{5 mm} To separate eqns.~(\ref{eq:expdirac1.c}-\ref{eq:expdirac4.c})
or (\ref{eq:expdirac1.k}-\ref{eq:expdirac4.k}),
we assign the standard \mbox{time- and azimuthal} dependence to $\Psi$,
with the
\mbox{above-mentioned} gauge transformation across the equiatorial plane:
\begin{equation}
\Psi = e^{-i \omega t} e^{i(m \mp eQ_{m}) \phi} \psi  \label{eq:septphi}
\end{equation}
\subsection{The Canonical Tetrad}

Substituting eqn.~(\ref{eq:septphi}) into
(\ref{eq:expdirac1.c}-\ref{eq:expdirac4.c}) gives, after some algebra,
\begin{subequations}
\begin{eqnarray}
- \frac{\delta}{\rho} ({\cal D}_{+}
                  + \frac{1}{2 \bar{\rho}^{*}}) \psi_{2}
+ \frac{i}{\rho} ({\cal L}_{-}
        + \frac{i a \sin\! \theta}{2 \bar{\rho}^{*}} ) \psi_{3}
                    & = & - i \mu \psi_{0} \label{eq:sep0.c}  \\
- \frac{i}{\rho} ({\cal L}_{+}
        + \frac{i a \sin\! \theta}{2 \bar{\rho}^{*}} ) \psi_{2}
+ \frac{\delta}{\rho} ({\cal D}_{-}
                    + \frac{1}{2 \bar{\rho}^{*}} ) \psi_{3}
                    & = & - i \mu \psi_{1} \label{eq:sep1.c}  \\
\frac{\delta}{\rho} ({\cal D}_{-}
           + \frac{1}{2 \bar{\rho}} ) \psi_{0}
- \frac{i}{\rho} ({\cal L}_{-}
            - \frac{i a \sin\! \theta}{2 \bar{\rho}} ) \psi_{1}
                   & = & - i \mu \psi_{2} \label{eq:sep2.c}  \\
\frac{i}{\rho} ({\cal L}_{+}
        - \frac{i a \sin\! \theta}{2 \bar{\rho}} ) \psi_{0}
- \frac{\delta}{\rho} ({\cal D}_{+}
                   + \frac{1}{2 \bar{\rho}} ) \psi_{1}
                     & = & - i \mu \psi_{3} \label{eq:sep3.c}
\end{eqnarray}
\end{subequations}
where ${\cal D}_{\pm}^{(C)}$ and ${\cal L}_{\pm}^{(C)}$
are purely radial and purely angular, respectively:
\begin{subequations}
\begin{eqnarray}
{\cal D}_{+}^{(C)} = \partial_{r}
                      + i \frac{(r^{2}+a^{2})\omega + e Q_{e} r - m a}{\Delta}
                  + \frac{(r-M)}{2 \Delta}      \\
{\cal D}_{-}^{(C)} = \partial_{r}
                      - i \frac{(r^{2}+a^{2})\omega + e Q_{e} r - m a}{\Delta}
                  + \frac{(r-M)}{2 \Delta}
\end{eqnarray}
\end{subequations}
\begin{subequations}
\begin{eqnarray}
{\cal L}_{+}^{(C)} = \partial_{\theta} - a \omega \sin\! \theta
            + \frac{m}{\sin\! \theta} + (\frac{1}{2} + e Q_{m}) \cot \theta \\
{\cal L}_{-}^{(C)} = \partial_{\theta} + a \omega \sin\! \theta
            - \frac{m}{\sin\! \theta} + (\frac{1}{2} - e Q_{m}) \cot \theta
\end{eqnarray}
\end{subequations}
Defining $\psi_{0} = f_{0}/\sqrt{\bar{\rho}} $,
$\psi_{1} = f_{1}/\sqrt{\bar{\rho}} $,
$\psi_{2} = f_{2}/\sqrt{\bar{\rho}^{*}} $,
and $ \psi_{3} = f_{3}/\sqrt{\bar{\rho}^{*}} $;
multiplying eqns. (\ref{eq:sep0.c}) and (\ref{eq:sep1.c}) by
$\rho\sqrt{\bar{\rho}^{*}}$, eqns.~(\ref{eq:sep2.c}) and (\ref{eq:sep3.c}) by
$\rho\sqrt{\bar{\rho}}$, we get
\begin{subequations}
\begin{eqnarray}
- \delta \; {\cal D}_{+} f_{2} + i {\cal L}_{-} f_{3}
                   = - i \mu (r - i a \cos\! \theta) f_{0} \label{eq:sep10.c}
\\
\delta \; {\cal D}_{-} f_{3} - i {\cal L}_{+}  f_{2}
                   = - i \mu (r - i a \cos\! \theta) f_{1} \label{eq:sep11.c}
\\
\delta \; {\cal D}_{-} f_{0} - i {\cal L}_{-} f_{1}
                   = - i \mu (r + i a \cos\! \theta) f_{2} \label{eq:sep12.c}
\\
- \delta \; {\cal D}_{+} f_{1} + i {\cal L}_{+} f_{0}
                    = - i \mu (r + i a \cos\! \theta) f_{3} \label{eq:sep13.c}
\end{eqnarray}
\end{subequations}
The structure of eqns.(\ref{eq:sep10.c}-\ref{eq:sep13.c}) suggests that they
can
be separated by the substitutions
\begin{equation}
\begin{array}{ccc}
f_{0} = R_{-} S_{+}
& ; &  f_{1} = R_{+} S_{-} \\
f_{2} = R_{+} S_{+}
& ; &  f_{3} = R_{-} S_{-}
\end{array}
\end{equation}
which gives
\begin{subequations}
\begin{eqnarray}
\delta \; {\cal D}_{+} R_{+} - i \mu r  R_{-}
                       = \lambda_{0} R_{-} & ; &
i {\cal L}_{-} S_{-} + \mu a \cos\! \theta S_{+}
                     = \lambda_{0} S_{+}          \label{eq:sepr0.c}  \\
\delta \; {\cal D}_{-} R_{-} + i \mu r  R_{+}
                       = \lambda_{1} R_{+} & ; &
i {\cal L}_{+} S_{+} - \mu a \cos\! \theta S_{-}
                     = \lambda_{1} S_{-}           \label{eq:sepr1.c}  \\
\delta \; {\cal D}_{-} R_{-} + i \mu r  R_{+}
                       = \lambda_{2} R_{+} & ; &
i {\cal L}_{-} S_{-} + \mu a \cos\! \theta S_{+}
                     = \lambda_{2} S_{+}           \label{eq:sepr2.c}  \\
\delta \; {\cal D}_{+} R_{+} - i \mu r  R_{-}
                       = \lambda_{3} R_{-} & ; &
i {\cal L}_{+} S_{+} - \mu a \cos\! \theta S_{-}
                     = \lambda_{3} S_{-}             \label{eq:sepr3.c}
\end{eqnarray}
\end{subequations}
The consistency of these equations requires
\begin{equation}
\lambda_{0} = \lambda_{3} = \lambda_{1} = \lambda_{2}
= \lambda
\end{equation}
and we are left with
\begin{subequations}
\begin{eqnarray}
\delta \; {\cal D}_{+}^{(C)} R_{+}^{(C)} - i \mu r  R_{-}^{(C)}
                          = \lambda R_{-}^{(C)}  \label{eq:rminus.c}  \\
\delta \; {\cal D}_{-}^{(C)} R_{-}^{(C)} + i \mu r  R_{+}^{(C)}
                       = \lambda R_{+}^{(C)}       \label{eq:rplus.c}
\end{eqnarray}
\end{subequations}
and
\begin{subequations}
\begin{eqnarray}
{\cal L}_{+}^{(C)} S_{+}^{(C)} + i \mu a \cos\! \theta S_{-}^{(C)}
                  = - i \lambda S_{-}^{(C)}        \label{eq:splus.c}  \\
{\cal L}_{-}^{(C)} S_{-}^{(C)} - i \mu a \cos\! \theta S_{+}^{(C)}
                     = - i \lambda S_{+}^{(C)}       \label{eq:sminus.c}
\end{eqnarray}
\end{subequations}
Or, by combining pairs, we can get decoupled $2^{{\rm nd}}$ order equations
\begin{subequations}
\begin{eqnarray}
{[ \delta \; {\cal D}_{-}^{(C)} \delta \; {\cal D}_{+}^{(C)}
- \frac{i \mu \Delta}{(\lambda+i \mu r)} {\cal D}_{+}^{(C)}
- (\lambda^{2} + \mu^{2} r^{2} ) ]} R_{+}^{(C)} = 0   \label{eq:rplus2.c}  \\
{[ \delta \; {\cal D}_{+}^{(C)} \delta \; {\cal D}_{-}^{(C)}
+ \frac{i \mu \Delta}{(\lambda-i \mu r)} {\cal D}_{-}^{(C)}
- (\lambda^{2} + \mu^{2} r^{2} ) ]} R_{-}^{(C)} = 0      \label{eq:rminus2.c}
\end{eqnarray}
\end{subequations}
\begin{subequations}
\begin{eqnarray}
{[ {\cal L}_{-}^{(C)} {\cal L}_{+}^{(C)}
+ \frac{\mu a \sin\! \theta}
{\lambda + \mu a \sin\! \theta }  {\cal L}_{+}^{(C)}
+ (\lambda^{2} - \mu^{2} a^{2} \cos^{2}\! \theta) ]} S_{+}^{(C)} = 0
                                                         \label{eq:splus2.c} \\
{[ {\cal L}_{+}^{(C)} {\cal L}_{-}^{(C)}
- \frac{\mu a \sin\! \theta}
{\lambda - \mu a \cos\! \theta}   {\cal L}_{-}^{(C)}
+ (\lambda^{2} - \mu^{2} a^{2} \cos^{2}\! \theta) ]} S_{-}^{(C)}  = 0
                                                     \label{eq:sminus2.c}
\end{eqnarray}
\end{subequations}

\subsection{The Kinnersley-type Tetrad}
\hspace{5 mm}
The same \mbox{time- and azimuthal} dependence (\ref{eq:septphi}), when
substituted into eqns.~(\ref{eq:expdirac1.k}-\ref{eq:expdirac4.k}) gives
\begin{subequations}
\begin{eqnarray}
\frac{\Delta}{\sqrt{2}\, \rho^{2}} {\cal D}_{+} \psi_{2}
+ \frac{1}{\bar{\rho}} ({\cal L}_{+}
          + \frac{i a \sin\! \theta}{\bar{\rho}^{*}} ) \psi_{3}
                & = & i \mu \psi_{0} \label{eq:sep0.k}  \\
\sqrt{2} ( {\cal D}_{-} + \frac{1}{\bar{\rho}^{*}} ) \psi_{3}
             - \frac{1}{\bar{\rho}^{*}} {\cal L}_{-}  \psi_{2}
                & = & - i \mu \psi_{1} \label{eq:sep1.k}  \\
\sqrt{2} ( {\cal D}_{-} + \frac{1}{\bar{\rho}} ) \psi_{0}
         + \frac{1}{\bar{\rho}} {\cal L}_{+} \psi_{1}
                & = & - i \mu \psi_{2} \label{eq:sep2.k}  \\
\frac{\Delta}{\sqrt{2}\, \rho^{2}} {\cal D}_{+} \psi_{1}
     - \frac{1}{\bar{\rho}^{*}}
         ( {\cal L}_{-} - \frac{i a \sin\! \theta}{\bar{\rho}} )
        \psi_{0}  & = & i \mu \psi_{3}   \label{eq:sep3.k}
\end{eqnarray}
\end{subequations}
the radial and angular operators ${\cal D}_{\pm}^{(K)}$ and
${\cal L}_{\pm}^{(K)}$ being
\begin{subequations}
\begin{eqnarray}
{\cal D}_{+}^{(K)} & = & \partial_{r} + \frac{i}{\Delta}
                 [(r^{2}+a^{2})\omega - m a + e Q_{e} r]
                  + \frac{(r-M)}{\Delta}      \\
{\cal D}_{-}^{(K)} & = & \partial_{r} - \frac{i}{\Delta}
                 [(r^{2}+a^{2})\omega - m a + e Q_{e} r]
\end{eqnarray}
\end{subequations}
\begin{subequations}
\begin{eqnarray}
{\cal L}_{+}^{(K)} & = & \partial_{\theta} + a \omega \sin\! \theta
          - \frac{m}{\sin\! \theta}
          + (\frac{1}{2} + e Q_{m}) \cot \theta \\
{\cal L}_{-}^{(K)} & = & \partial_{\theta} - a \omega \sin\! \theta
          + \frac{m}{\sin\! \theta}
          + (\frac{1}{2} - e Q_{m}) \cot \theta
\end{eqnarray}
\end{subequations}
Multiplying eqns. (\ref{eq:sep0.k}) and (\ref{eq:sep3.k}) by
$\rho^{2} = \bar{\rho} \bar{\rho}^{*}$, eq.~(\ref{eq:sep1.k}) by
$\bar{\rho}^{*} $, eq.~(\ref{eq:sep2.k}) by $\bar{\rho}$; defining
$F_{0} = \bar{\rho} \psi_{0}$, $F_{1} = \psi_{1}$,
$F_{2} = \psi_{2}$, and $F_{3} = \bar{\rho}^{*} \psi_{3}$,
we get
\begin{subequations}
\begin{eqnarray}
\frac{\Delta}{\sqrt{2}} {\cal D}_{+} F_{2} + {\cal L}_{+} F_{3}
               & = & i \mu (r - i a \cos\! \theta) F_{0} \label{eq:sep10.k} \\
\sqrt{2} {\cal D}_{-} F_{3} - {\cal L}_{-}  F_{2}
               & = & - i \mu (r - i a \cos\! \theta) F_{1} \label{eq:sep11.k}
\\
\sqrt{2} {\cal D}_{-} F_{0} + {\cal L}_{+} F_{1}
               & = & - i \mu (r + i a \cos\! \theta) F_{2} \label{eq:sep12.k}
\\
\frac{\Delta}{\sqrt{2}} {\cal D}_{+} F_{1} - {\cal L}_{-} F_{0}
               & = & i \mu (r + i a \cos\! \theta) F_{3} \label{eq:sep13.k}
\end{eqnarray}
\end{subequations}
Again, the eqns.(\ref{eq:sep10.k}-\ref{eq:sep13.k}) can be separated by
\begin{equation}
\begin{array}{ccc}
F_{0} = R_{-} S_{+} & ; &  F_{1} = R_{+} S_{-} \\
F_{2} = R_{+} S_{+} & ; &  F_{3} = R_{-} S_{-}
\end{array}
\end{equation}
which gives
\begin{subequations}
\begin{eqnarray}
\frac{\Delta}{\sqrt{2}} {\cal D}_{+} R_{+} - i \mu r  R_{-}
                 = \lambda_{0} R_{-} & ; &
{\cal L}_{+} S_{-} - \mu a \cos\! \theta S_{+}
                 = - \lambda_{0} S_{+}        \label{eq:sepr0.k}  \\
\sqrt{2} {\cal D}_{-} R_{-} + i \mu r  R_{+}
                 = \lambda_{1} R_{+} & ; &
{\cal L}_{-} S_{+} - \mu a \cos\! \theta S_{-}
                 = \lambda_{1} S_{-}          \label{eq:sepr1.k}  \\
\sqrt{2} {\cal D}_{-} R_{-} + i \mu r  R_{+}
                 = \lambda_{2} R_{+} & ; &
{\cal L}_{+} S_{-} - \mu a \cos\! \theta S_{+}
                 = - \lambda_{2} S_{+}        \label{eq:sepr2.k}  \\
\frac{\Delta}{\sqrt{2}} {\cal D}_{+} R_{+} - i \mu r  R_{-}
                 = \lambda_{3} R_{-} & ; &
{\cal L}_{-} S_{+} - \mu a \cos\! \theta S_{-}
                 = \lambda_{3} S_{-}           \label{eq:sepr3.k}
\end{eqnarray}
\end{subequations}
as before,
\begin{equation}
\lambda_{0} = \lambda_{3} = \lambda_{1} = \lambda_{2}
= \lambda '
\end{equation}
therefore
\begin{subequations}
\begin{eqnarray}
\frac{\Delta}{\sqrt{2}} {\cal D}_{+}^{(K)} R_{+}^{(K)} - i \mu r  R_{-}^{(K)}
                       = \lambda ' R_{-}^{(K)}   \label{eq:rminus.k}  \\
\sqrt{2} {\cal D}_{-}^{(K)} R_{-}^{(K)} + i \mu r  R_{+}^{(K)}
                       = \lambda ' R_{+}^{(K)}   \label{eq:rplus.k}
\end{eqnarray}
\end{subequations}
\begin{subequations}
\begin{eqnarray}
{\cal L}_{+}^{(K)} S_{-}^{(K)} - \mu a \cos\! \theta S_{+}^{(K)}
                     = - \lambda ' S_{+}^{(K)}    \label{eq:splus.k}  \\
{\cal L}_{-}^{(K)} S_{+}^{(K)} - \mu a \cos\! \theta S_{-}^{(K)}
                     = \lambda ' S_{-}^{(K)}      \label{eq:sminus.k}
\end{eqnarray}
\end{subequations}
The corresponding decoupled $2^{{\rm nd}}$ order equations are
\begin{subequations}
\begin{eqnarray}
{[ \Delta {\cal D}_{+}^{(K)} {\cal D}_{-}^{(K)}
+ \frac{i \mu \Delta}{(\lambda '-i \mu r)} {\cal D}_{-}^{(K)}
- (\lambda '^{2} + \mu^{2} r^{2} ) ]} R_{-}^{(K)} = 0  \label{eq:rminus2.k}  \\
{[ {\cal D}_{-}^{(K)} \Delta {\cal D}_{+}^{(K)}
- \frac{i \mu}{(\lambda '+i \mu r)} \Delta {\cal D}_{+}^{(K)}
- (\lambda '^{2} + \mu^{2} r^{2} ) ]} R_{+}^{(K)} = 0      \label{eq:rplus2.k}
\end{eqnarray}
\end{subequations}
\begin{subequations}
\begin{eqnarray}
{[ {\cal L}_{+}^{(K)} {\cal L}_{-}^{(K)}
+ \frac{\mu a \sin\! \theta}
{\lambda ' + \mu a \cos\! \theta}  {\cal L}_{-}^{(K)}
+ (\lambda '^{2} - \mu^{2} a^{2} \cos^{2}\! \theta) ]}  S_{+}^{(K)}  = 0
                                                     \label{eq:splus2.k}  \\
{[ {\cal L}_{-}^{(K)} {\cal L}_{+}^{(K)}
- \frac{\mu a \sin\! \theta}
{\lambda ' - \mu a \cos\! \theta }{\cal L}_{+}^{(K)}
+ (\lambda '^{2} - \mu^{2} a^{2} \cos^{2}\! \theta) ]}  S_{-}^{(K)} = 0
                                                          \label{eq:sminus2.k}
\end{eqnarray}
\end{subequations}

\section{Comparison with previous results}
\hspace{5 mm}
Both papers that we compare our results against use the \mbox{Newman-Penrose}
formalism. For the simpler `canonical tetrad', our results
(eqns.\ref{eq:rminus.c}-\ref{eq:sminus2.c})
agree with Carter \& McLenaghan \cite{cart-mcl.MG2} in the
limit $Q_{m} = 0$, with
${\cal D}_{\pm}^{(C)} \rightarrow {\cal D}_{\pm\frac{1}{2}}$,
${\cal L}_{\pm}^{(C)}  \rightarrow {\cal L}_{\pm\frac{1}{2}}$,
$R_{\pm}^{(C)}  \rightarrow X_{\pm\frac{1}{2}}$,
$S_{+}^{(C)}  \rightarrow Y_{\frac{1}{2}}$,
$S_{-}^{(C)}  \rightarrow -Y_{-\frac{1}{2}}$,
$\lambda \rightarrow \sqrt{2}\,\lambda$,
$\mu \rightarrow \sqrt{2}\, \mu$.
For the \mbox{Kinnersley-type} tetrad, the equations
(\ref{eq:rminus.k}-\ref{eq:sminus2.k}) agree with
Chandrasekhar \cite{chandra.book} in the limit $e=Q_{e}=Q_{m}=0$, with
${\cal D}_{+}^{(K)} \rightarrow {\cal D}_{\frac{1}{2}}^{\dag}$,
${\cal D}_{-}^{(K)} \rightarrow {\cal D}_{0}$,
${\cal L}_{+}^{(K)} \rightarrow {\cal L}_{\frac{1}{2}}^{\dag}$,
${\cal L}_{-}^{(K)}\rightarrow {\cal L}_{\frac{1}{2}}$,
$S_{\pm}^{(K)} \rightarrow S_{\pm \frac{1}{2}}$,
$R_{+}^{(K)} \rightarrow R_{+\frac{1}{2}}$,
$R_{-}^{(K)} \rightarrow - R_{-\frac{1}{2}}$,
$\lambda ' \rightarrow - \lambda$.
A difference with the $Q_{m} = 0$ case is that $m$ can now also take on
\mbox{half-integer} values: To make the wavefunctions
(\ref{eq:septphi}) \mbox{single-valued},
both $m + eQ_{m}$ and $m-eQ_{m}$ have to be integers, which makes $m$ and
$eQ_{m}$ integers or \mbox{half-integers}, the latter requirement being,
of course, the \mbox{well-known} Dirac quantization condition.
Since the explicit forms of equations (\ref{eq:rplus2.c}-\ref{eq:sminus2.c})
and (\ref{eq:rminus2.k}-\ref{eq:sminus2.k})
are not very illuminating, we leave them in their present form.
Unfortunately, the angular equations (\ref{eq:splus2.c},\ref{eq:sminus2.c},
\ref{eq:splus2.k},\ref{eq:sminus2.k}), unlike in the simpler scalar case
\cite{semiz2}, are not of the \mbox{Sturm-Liouville} form, therefore we cannot
make a statement about the completeness of the solutions.

The Chandrasekhar result has been also generalized to other
cases \cite{guven},\cite{rudiger},\cite{kamr-mcl.84},\cite{iyer-kamr},
the mathematical structure of the problem has been investigated
\cite{cart-mcl.79},\cite{cart-mcl.MG2},\cite{mcl-spindel},\cite{kamr-mcl.83},
\cite{rudiger},\cite{kamr-mcl.84},\cite{iyer-vish.85},\cite{iyer-vish.87}
and the solutions studied
\cite{suffern.et.al},\cite{chakra},\cite{kalnins-miller}.
Other approaches to the problem of separation of variables in the Dirac
Equation on curved spaces include the St\"{a}ckel Space method \cite{stackel}
and the "algebraic" method \cite{alg}.

\vspace{.5cm}

{\bf Acknowledgements :}\\

I am grateful to V. Moncrief for valuable discussions and helpful
suggestions. This research was partially supported by NSF grant
PHY-8903939
to Yale University.


\begin{thebibliography}{50}

\bibitem{brill-whe} D. Brill and J. A. Wheeler,
                     Rev. Mod. Phys. {\bf 29} 645 (1957).

\bibitem{teukolsky} S. A. Teukolsky,
                                Astrophysical J. {\bf 185}, 635 (1973).

\bibitem{unruh} W. Unruh, Phys. Rev. Lett. {\bf 31}, 1265 (1973)

\bibitem{chandra.paper} S. Chandrasekhar, Proc. R. Soc. {\bf A 349}, 571
(1976);
                        see S. Chandrasekhar, Proc. R. Soc. {\bf A 350},
                        564 (1976) for errata; also see \cite{chandra.book},
                        \S 104.

\bibitem{page} D. N. Page, Phys. Rev. {\bf D 14}, 1509, (1976).

\bibitem{toop} N. Toop, preprint, D.A.M.T.P., Cambridge (1976).

\bibitem{plebanski-dem}  J. F. Plebanski and M. Demianski,
                              Ann. Phys. {\bf 98}, 98 (1976).

\bibitem{dudley-finley} A. L. Dudley and J. D. Finley,
                                   J.Math. Phys. {\bf 20}, 311 (1979).

\bibitem{semiz1} \.{I}. Semiz,  Class. Quant. Grav. {\bf 7}, 353 (1990).

\bibitem{wbg} S. Weinberg, {\em Gravitation and Cosmology}, Wiley, New York
                                 (1972).

\bibitem{bir-dav} N.D. Birrell and P.C.W. Davies, {\em Quantum Fields in Curved
                                Space}, Cambridge (1982).

\bibitem{chandra.book} S. Chandrasekhar, {\em The Mathematical Theory of Black
                                           Holes}, Oxford (1983).

\bibitem{kasuya} M. Kasuya, Phys. Lett. {\bf 103B}, 351 (1981).

\bibitem{wu-yang} T. T. Wu and C. N. Yang,
                                   Nucl. Phys. {\bf B107}, 365 (1976).

\bibitem{tetrad.m} J. M. Aguirregabiria,
                      The Mathematica Journal, {\bf v.1}, issue 2, 51 (1990).

\bibitem{semiz2} \.{I}. Semiz, Phys. Rev. {\bf D 45}, 532 (1992).

\bibitem{guven} R. G\"{u}ven, Proc. R. Soc. {\bf A 356}, 465 (1977).

\bibitem{cart-mcl.79} B. Carter and R.G McLenaghan,
                                          Phys. Rev. D {\bf 19}, 1093 (1979).

\bibitem{cart-mcl.MG2} B. Carter and R.G McLenaghan, pp.575 of
                       {\em Proceedings of the Second Marcel Grossman Meeting
                       on General Relativity, 1979}, ed. R. Ruffini,
                       North Holland (1982)

\bibitem{mcl-spindel} R.G McLenaghan and Ph. Spindel,
      Phys. Rev. D {\bf 20}, 409 (1979); Bull. Soc. Math. Belg. XXXI, 65
(1979).

\bibitem{kamr-mcl.83} N. Kamran and R.G McLenaghan,
                                        Lett. Math. Phys. {\bf 7}, 381 (1983).

\bibitem{rudiger} R. R\"{u}diger, J. Math. Phys. {\bf 25}, 649 (1984).

\bibitem{kamr-mcl.84} N. Kamran and R.G McLenaghan,
                                       J. Math. Phys. {\bf 25}, 1019 (1984).

\bibitem{iyer-vish.85} B.R. Iyer and C. V. Vishveshwara,
                                   J. Math. Phys. {\bf 26}, 1034 (1985).

\bibitem{iyer-vish.87} B.R. Iyer and C. V. Vishveshwara,
                                      J. Math. Phys. {\bf 28}, 1377 (1987).

\bibitem{suffern.et.al} K. G. Suffern, E. D. Fackarell and C.M. Cosgrove,
                                        J. Math. Phys. {\bf 24}, 1350 (1983).

\bibitem{chakra} S. K. Chakrabarti, Proc. R. Soc. {\bf A 391}, 27 (1984).

\bibitem{kalnins-miller} E. G. Kalnins and W. Miller,
                                        J. Math. Phys. {\bf 33}, 286 (1992).

\bibitem{iyer-kamr} B.R. Iyer and N. Kamran,
                                      J. Math. Phys. {\bf 32}, 2497 (1991).

\bibitem{stackel} V. G. Bagrov, A. V. Shapovalov and A. A. Yevseyevich,
                  Class. Quantum Gravity {\bf 8}, 163 (1991);
                  and references therein.

\bibitem{alg}  G. V. Shishkin and W. D. Cabos,
               J. Math. Phys. {\bf 33}, 916 (1992); and references therein.

\end{thebibliography}
\end{document}